%% file: main.tex
\documentclass[a4paper,11pt]{article}
\usepackage{pos}

\usepackage{array} 
\usepackage{bm} 
\usepackage{hyperref}
\usepackage{multirow}
\usepackage{xcolor}
\usepackage{subfigure}
\usepackage{adjustbox}
\usepackage{enumitem}
\usepackage{caption}
\usepackage{anyfontsize}
\usepackage{color}
\usepackage{fontawesome5}
\usepackage{orcidlink}

\graphicspath{{./figs/}}

\allowdisplaybreaks

\input{macro.tex}

\title{Machine Learning-Based Estimation of Cumulants of Chiral
  Condensate via Multi-Ensemble Reweighting with
  \raisebox{-0.13em}{\includegraphics[height=1em]{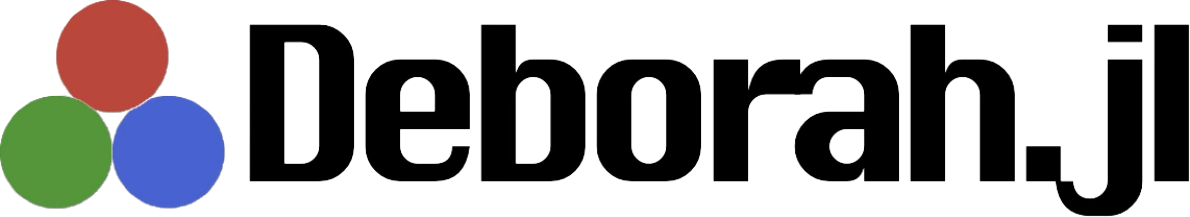}}}
\ShortTitle{ML Estimation of Cumulants}

\author*[a]{Benjamin J. Choi\,\orcidlink{\orcidauthorCHOI}}
\author[a]{Hiroshi Ohno\,\orcidlink{\orcidauthorOHNO}}
\author[b,c,d]{Akio Tomiya\,\orcidlink{\orcidauthorTOMIYA}}

\affiliation[a]{Center for Computational Sciences, 
University of Tsukuba,\\
1-1-1 Tennodai, Tsukuba, Ibaraki 305-8577, Japan}

\affiliation[b]{Department of Information and Mathematical Sciences,
Tokyo Woman’s Christian University,\\
2-6-1 Zempukuji, Suginami-ku, Tokyo 167-8585, Japan}

\affiliation[c]{RIKEN Center for Computational Science, \\ 7-1-26
Minatojima-minami-machi, Chuo-ku, Kobe 650-0047, Japan}

\affiliation[d]{Department of Physics, Kyoto University, Kitashirakawa,
Sakyo-ku, Kyoto 606-8502, Japan}

\emailAdd{benchoi@ccs.tsukuba.ac.jp}
\emailAdd{hohno@ccs.tsukuba.ac.jp}
\emailAdd{akio@yukawa.kyoto-u.ac.jp}

\abstract{We investigate a bias-corrected machine learning (ML)
  strategy for estimating traces of the inverse Dirac operator, $\Tr\,
  M^{-n}$ ($n=1,2,3,4$), motivated by the need for higher-order
  cumulants of the chiral condensate near the finite-temperature QCD
  critical endpoint.
  Our supervised regression framework is trained on Wilson-clover
  ensembles with the Iwasaki gauge action, and we explore two input
  feature scenarios: one using $\Tr\, M^{-1}$ and another relying
  solely on gauge observables (plaquette and rectangle), enabling a
  fully feature-based prediction pipeline.
  Using $\Tr\, M^{-1}$ both as a physical input to cumulant
  construction and as a feature for predicting higher powers, we find
  that even with $\sim1\%$ labeled data, the resulting susceptibility,
  skewness, and kurtosis remain statistically consistent with fully
  measured baselines, reducing computational cost to about $26\%$.
  In the feature-only approach, where correlations rather than
  explicit stochastic traces drive the predictions, bias correction
  plays a more pronounced role.
  We quantify this impact through multi ensemble reweighting across
  nearby quark masses.
  Our results demonstrate that bias-corrected ML estimates can
  significantly reduce measurement overhead while preserving the
  stability of higher-order observables relevant for locating the QCD
  critical endpoint.
  Code for this work is available at
  \href{https://github.com/saintbenjamin/Deborah.jl}{
    \faGithub~\raisebox{-0.13em}{\includegraphics[height=1em]{deborah_JuliaQCD}}}.
}

\FullConference{The 42nd International Symposium on Lattice Field
  Theory (LATTICE2025)\\
  2-8 November 2025\\
  Tata Institute of Fundamental Research, Mumbai, India\\}

\begin{document}

\maketitle

\section{Introduction}
\label{sec:intro}

Understanding the critical endpoint in the finite-temperature QCD
phase diagram requires the precise determination of higher-order
fluctuations of the chiral condensate, which acts as the order
parameter~\cite{Philipsen:2021qji, Guenther:2020jwe}.
Techniques such as the kurtosis intersection method rely on these
fluctuations to identify the critical behavior and extract universal
scaling properties near the transition~\cite{Jin:2014hea,
  Kuramashi:2016kpb}.

The evaluation of these fluctuations involves computing traces of
powers of the inverse Dirac operator, $\Tr \, M^{-n}$.
Even when stochastic estimators such as Hutchinson-based approaches
are used~\cite{Dong:1993pk}, the dominant computational effort comes
from repeated solutions of large sparse linear systems with iterative
solvers.
As a result, brute-force measurement at high statistics remains
costly.

Recent developments in artificial intelligence for scientific
applications have opened new possibilities for accelerating
computationally intensive studies in lattice field theory
\cite{Tomiya:2025quf}.
In particular, machine-learning-assisted approaches have begun to be
explored as a means to reduce the cost of stochastic measurements
while maintaining physics fidelity.

A recent strategy to mitigate this computational burden applies
supervised machine learning together with a bias correction scheme
motivated by the All Mode Averaging (AMA) framework~\cite{Bali:2009hu,
  Blum:2012uh}, as demonstrated in Ref.~\cite{Yoon:2018krb}.
In this approach, explicitly measured configurations form a labeled
set that is divided into training and bias-correction subsets, while
the remaining unlabeled configurations are used for prediction,
balancing model accuracy against bias control.

In this work, we systematically vary both the size of the labeled
dataset and the training fraction within it, assessing how these
choices influence the precision of ML-based estimators for $\Tr \,
M^{-n}$.
Beyond the direct trace estimates, we explore a physics application of
the ML outputs through multi-ensemble reweighting across nearby quark
masses.
These tests probe whether the ML-derived observables retain sufficient
fidelity for thermodynamic analyses relevant to the QCD critical
endpoint.

We further compare bias-corrected and uncorrected setups to isolate
the impact of bias removal.
Conventional full-statistics evaluations are carried out in parallel
to provide a direct baseline for judging the accuracy and stability of
the ML-assisted estimation workflow.
The implementation used in this study is provided within the
\href{https://github.com/saintbenjamin/Deborah.jl}{
  \faGithub~\raisebox{-0.13em}{\includegraphics[height=1em]{deborah_JuliaQCD}}}
framework \cite{Choi:2026zen}.

\vspace{-0.3em}

\section{Formalism, Dataset, and ML Estimation Framework}
\label{sec:formalism}

\subsection{Notation, Dataset, and Cumulant Observables}
\label{sec:cumulant-eval}

\begin{table}[tb]
  \vspace{-1.0em}
  \renewcommand{\arraystretch}{1.1}
  \centering
  \begin{adjustbox}{max width=\textwidth}
    \begin{tabular}{@{\quad}\.r@{\quad}|@{\quad}^l@{\quad}}
      \hline
      \hline
      Symbol & Description \\
      \hline
      $X$ & input observables (or feature) (\textit{e.g.},~$X =
      \text{Plaquette}$, $\text{Rectangle}$, $\Tr \, M^{-1}$) \\
      $Y$ & output observables (or target) (\textit{e.g.},~$Y = \Tr \,
      M^{-n}$ for $n=1,2,3,4$) \\
      $S^Z$ & the total dataset of $Z=X,Y$ where $S^Z =
      S^Z_{\text{LB}} \cup S^Z_{\text{UL}}$ \\
      $S^Z_{\text{LB}}$ & the labeled set of the original data for
      $Z=X,Y$ where $S^Z_{\text{LB}} = S^Z_{\text{TR}} \cup
      S^Z_{\text{BC}}$ \\
      $S^Z_{\text{TR}}$ & the training set of the original data for
      $Z=X,Y$ \\
      $S^Z_{\text{BC}}$ & the bias correction set of the original data
      for $Z=X,Y$ \\
      $S^Z_{\text{UL}}$ & the unlabeled set of the original data for
      $Z=X,Y$ \\
      $N$ & the number of elements of $S^Z$ where $N =
      \left|S^{X}\right| = \left|S^{Y}\right|$ \\
      $N_{\text{LB}}$ & the number of elements of $S^Z_{\text{LB}}$
      where $N_{\text{LB}} = \left|S^{X}_{\text{LB}}\right| =
      \left|S^{Y}_{\text{LB}}\right|$ \\
      $N_{\text{TR}}$ & the number of elements of $S^Z_{\text{TR}}$
      where $N_{\text{TR}} = \left|S^{X}_{\text{TR}}\right| =
      \left|S^{Y}_{\text{TR}}\right|$ \\
      $N_{\text{BC}}$ & the number of elements of $S^Z_{\text{BC}}$
      where $N_{\text{BC}} = \left|S^{X}_{\text{BC}}\right| =
      \left|S^{Y}_{\text{BC}}\right|$ \\
      $N_{\text{UL}}$ & the number of elements of $S^Z_{\text{UL}}$
      where $N_{\text{UL}} = \left|S^{X}_{\text{UL}}\right| =
      \left|S^{Y}_{\text{UL}}\right|$ \\
      $f(X)$ & the model trained with $S^{X}_{\text{TR}}$ and
      $S^{Y}_{\text{TR}}$ \\
      $Y^P$ & the ML estimation on $Y$ \\
      $S^P_{\text{BC}}$ & the bias correction set composed of the ML
      estimations $Y^P_{\text{BC}}$ corresponding to $S^Y_{\text{BC}}$
      \\
      $S^P_{\text{UL}}$ & the unlabeled set composed of the ML
      estimations $Y^P_{\text{UL}}$ corresponding to $S^Y_{\text{UL}}$
      \\
      \hline
      \hline
    \end{tabular}
  \end{adjustbox}
  \caption{Notation and convention used in this paper for the
    explanation of our work.}
  \label{tab:notation-1}
  \vspace{-0.3em}
\end{table}
%
For convenience, Table~\ref{tab:notation-1} provides the notation and
definitions that will be used throughout this paper.
These symbols will be referenced repeatedly in later sections, and the
table serves as a compact glossary rather than repeating explanations
inline.

\begin{table}[tb]
  \renewcommand{\arraystretch}{1.1} \centering
  \begin{adjustbox}{max width=\textwidth}
    \begin{tabular}{@{\qquad}c@{\quad}|@{\qquad}c@{\quad}|@{\qquad}c@{\quad}|@{\qquad}c@{\quad}|@{\qquad}c@{\quad}|@{\qquad}c@{\qquad}}
      \hline
      \hline
      \texttt{ID} & $N_{\text{S}}^3 \times N_{\text{T}}$ & $\beta$ &
      $c_{\text{SW}}$ & $\kappa$ & $N$ \\
      \hline
      \texttt{L12T4b1.60k13575} & $12^3 \times 4$ & 1.60 & 2.065 &
      0.13575 & 20000 \\
      \texttt{L12T4b1.60k13577} & $12^3 \times 4$ & 1.60 & 2.065 &
      0.13577 & 20000 \\
      \texttt{L12T4b1.60k13580} & $12^3 \times 4$ & 1.60 & 2.065 &
      0.13580 & 20000 \\
      \texttt{L12T4b1.60k13582} & $12^3 \times 4$ & 1.60 & 2.065 &
      0.13582 & 20000 \\
      \texttt{L12T4b1.60k13585} & $12^3 \times 4$ & 1.60 & 2.065 &
      0.13585 & 20000 \\
      \hline
      \hline
    \end{tabular}
  \end{adjustbox}
  \caption{Data used in this paper, originally produced for
    Ref.~\cite{Ohno:2018gcx}.
    Here, $N$ follows the same convention as in
    Table~\ref{tab:notation-1}, where it corresponds to the number of
    gauge configurations.}
  \label{tab:ensemble-1}
\end{table}
%
To test the machine learning (ML) estimation strategy in a practical
lattice-QCD environment, we utilize gauge configurations originally
produced for Ref.~\cite{Ohno:2018gcx}.
The ensembles were generated on the \texttt{Oakforest-PACS}
supercomputing system~\cite{Boku:2017urp} using the \texttt{BQCD}
framework~\cite{Nakamura:2010qh}, with $N_{\text{f}}=4$ Wilson–Clover
fermions~\cite{Sheikholeslami:1985ij} and the Iwasaki gauge
action~\cite{Iwasaki:1985we, Iwasaki:1983iya}.
This dataset serves as a well understood reference environment for
evaluating ML-based trace estimation performance.
The specific simulation parameters relevant to the present analysis
are listed in Table~\ref{tab:ensemble-1}.

We evaluate the impact of ML-based trace predictions on thermodynamic
observables derived from the Wilson-Clover operator,
\begin{align}
  M(x,y) &= \frac{1}{2\kappa}\,\delta_{x,y} +
  \frac{\mathrm{i}}{4}c_\textrm{sw} \,\sigma_{\mu\nu} F_{\mu\nu}(x) \,
  \delta_{x,y} - \frac{1}{2} \sum_{\mu=1}^{4}\sum_{s=\pm 1} \left( 1 -
  s \,\gamma_\mu \right) \, U_{s\,\mu}(x) \, \delta_{x,\,y +
    s\,\hat\mu} \,.
\end{align}
From the traces $\Tr\,M^{-n}$, we construct the quark-loop operators,
\begin{align}
  Q_1 &= N_\textrm{f} \; \Tr\,M^{-1} \,, \quad
  Q_2 = -N_\textrm{f} \; \Tr\,M^{-2} + (N_\textrm{f} \; \Tr\,M^{-1})^2 
  \,, \nonumber \\
  Q_3 &= 2N_\textrm{f} \; \Tr\,M^{-3} -3(N_\textrm{f} \;
  \Tr\,M^{-2})(N_\textrm{f} \; \Tr\,M^{-1}) + (N_\textrm{f} \;
  \Tr\,M^{-1})^3 \,, \nonumber \\
  Q_4 &= -6N_\textrm{f} \; \Tr\,M^{-4} +8(N_\textrm{f} \;
  \Tr\,M^{-3})(N_\textrm{f} \; \Tr\,M^{-1}) +3(N_\textrm{f} \;
  \Tr\,M^{-2})^2 \nonumber \\
  &\hphantom{=} -6(N_\textrm{f} \; \Tr\,M^{-2}) (N_\textrm{f} \;
  \Tr\,M^{-1})^2 + (N_\textrm{f} \; \Tr\,M^{-1})^4. \label{eq:Q_1234}
\end{align}
Using ensemble averages of $Q_i$, we obtain the cumulants of the
chiral condensate,
\begin{align}
  \Sigma &= C_1/V \,, \quad
  \chi = C_2/V \,, \quad
  S = C_3/C_2^{3/2} \,, \quad
  K = C_4/C_2^2 \,, \label{eq:cumulants-1}
\end{align}
with
\begin{align}
  C_1 &= \langle Q_1\rangle \,, \quad
  C_2 = \langle Q_2\rangle - \langle Q_1\rangle^2, \quad
  C_3 = \langle Q_3\rangle - 3\langle Q_2\rangle\langle Q_1\rangle +
  2\langle Q_1\rangle^3, \nonumber \\
  C_4 &= \langle Q_4\rangle -4\langle Q_3\rangle\langle Q_1\rangle
  -3\langle Q_2\rangle^2 + 12\langle Q_2\rangle\langle Q_1\rangle^2
  -6\langle Q_1\rangle^4. \label{eq:C_1234}
\end{align}

\subsection{Supervised Learning with Bias Correction}

In our supervised ML setup, each configuration carries input features
$X$ and target observables $Y$.
Only the labeled subset $S^Z_{\text{LB}}\subset S^Z$ ($Z=X,Y$) has
both $X$ and $Y$ available. We split this labeled set into a training
part $S^Z_{\text{TR}}$ and a bias-correction part $S^Z_{\text{BC}}$,
while the remaining configurations form the unlabeled set
$S^Z_{\text{UL}}$.
The basic two-step procedure is
\begin{enumerate}
\item Train a regression model $f(X)$ using $S^X_{\text{TR}}$ and
  $S^Y_{\text{TR}}$.
\item Apply $f$ to $X\in S^X_{\text{UL}}$ to obtain predictions $Y^P =
  f(X) \approx Y$.
\end{enumerate}
Because the model is trained on only a subset of the full data, it can
develop systematic biases.
Following the bias-correction strategy of Ref.~\cite{Yoon:2018krb}, we
use $S^Z_{\text{BC}}$ to correct this bias and construct the estimator
\begin{align}
  \bar{Y}_{\mathcal{P}1} &= \frac{1}{N_{\textrm{UL}}} \sum_{Y_i\in
    S^Y_{\text{UL}}} Y_i^{P} + \frac{1}{N_{\textrm{BC}}} \sum_{Y_j\in
    S^Y_{\text{BC}}}\bigl(Y_j - Y_j^{P}\bigr)\,, \label{eq:P1-1}
\end{align}
where $N_{\text{UL}}$ and $N_{\text{BC}}$ denote the sizes of
$S^Y_{\text{UL}}$ and $S^Y_{\text{BC}}$, respectively.
A second estimator, $\bar{Y}_{\mathcal{P}2}$, was also explored in
Ref.~\cite{Choi:2024ryu} but is not considered here; see that work for
its definition.

\subsection{Scanning the Labeled and Training Fractions}

To control the amount of exact CG work and the size of the
bias-correction sample, we parametrize the partition in terms of the
fractions
\begin{align}
  \mathcal{R}_{\text{LB}} &\equiv \frac{N_{\text{LB}}}{N}\,, \qquad
  \mathcal{R}_{\text{TR}} \equiv
  \frac{N_{\text{TR}}}{N_{\text{LB}}}\,,
\label{eq:R-LB-R-TR-1}
\end{align}
with $N = N_{\text{LB}} + N_{\text{UL}}$ and $N_{\text{LB}} =
N_{\text{TR}} + N_{\text{BC}}$.
We explore a set of representative values, $\mathcal{R}_{\text{LB}}
\in \{1,2,\dots,24,25\%\}$ and $\mathcal{R}_{\text{TR}} \in
\{0,10,\dots,90,100\%\}$, in order to map out how the ML estimation
quality depends on the amount of labeled data and on the training
fraction.

The extreme choices $\mathcal{R}_{\text{TR}} = 0\%$ and
$\mathcal{R}_{\text{TR}} = 100\%$ provide useful benchmarks.
In the former case, all labeled configurations are reserved for a
purely ``conventional'' estimate without ML, allowing us to monitor
convergence as $\mathcal{R}_{\text{LB}}$ increases.
In the latter case, all labeled data are used for training and no bias
correction is applied, so the impact of omitting the correction can be
directly assessed.

\subsection{Correlation Structure among Observables}

\begin{figure}[tb]
  \vspace{-1.5em}
  \subfigure[$\kappa = 0.13575$, \texttt{L12T4b1.60k13575} (the
    heaviest quark)]{
    \includegraphics[width=0.48\linewidth]{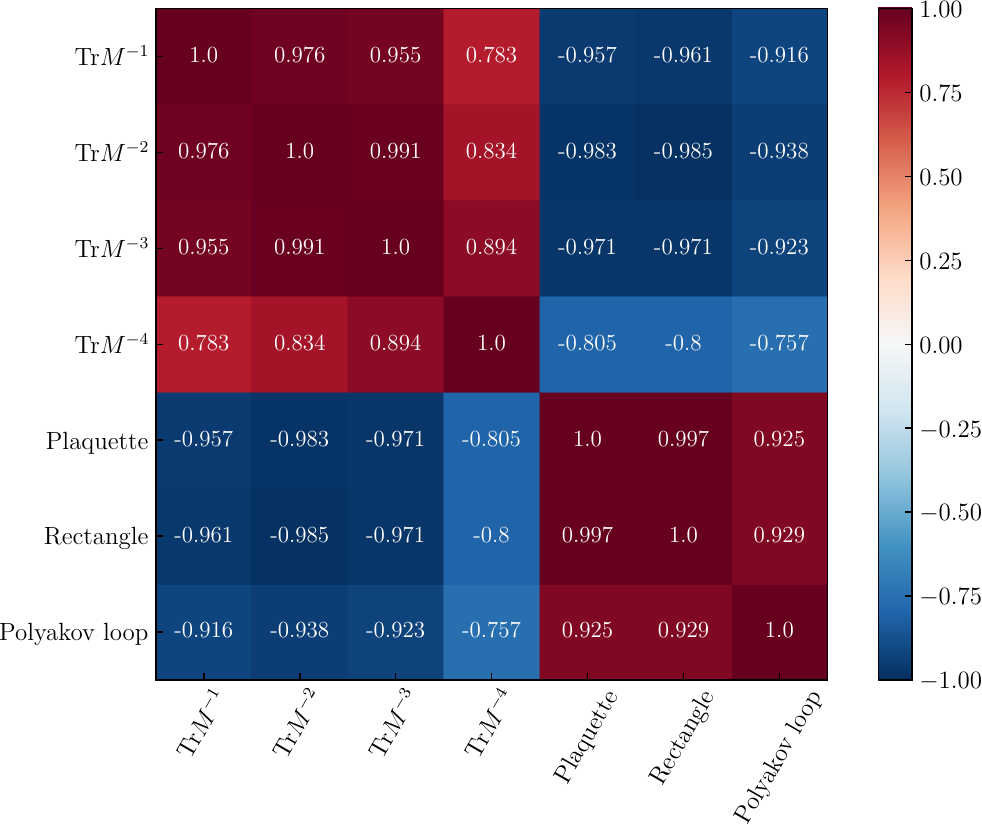}
        \label{fig:subfig-corr-ID-L12T4b1.60k13575}
  }
  \hfill
  \subfigure[$\kappa = 0.13590$, \texttt{L12T4b1.60k13585} (the
    lightest quark)]{
    \includegraphics[width=0.48\linewidth]{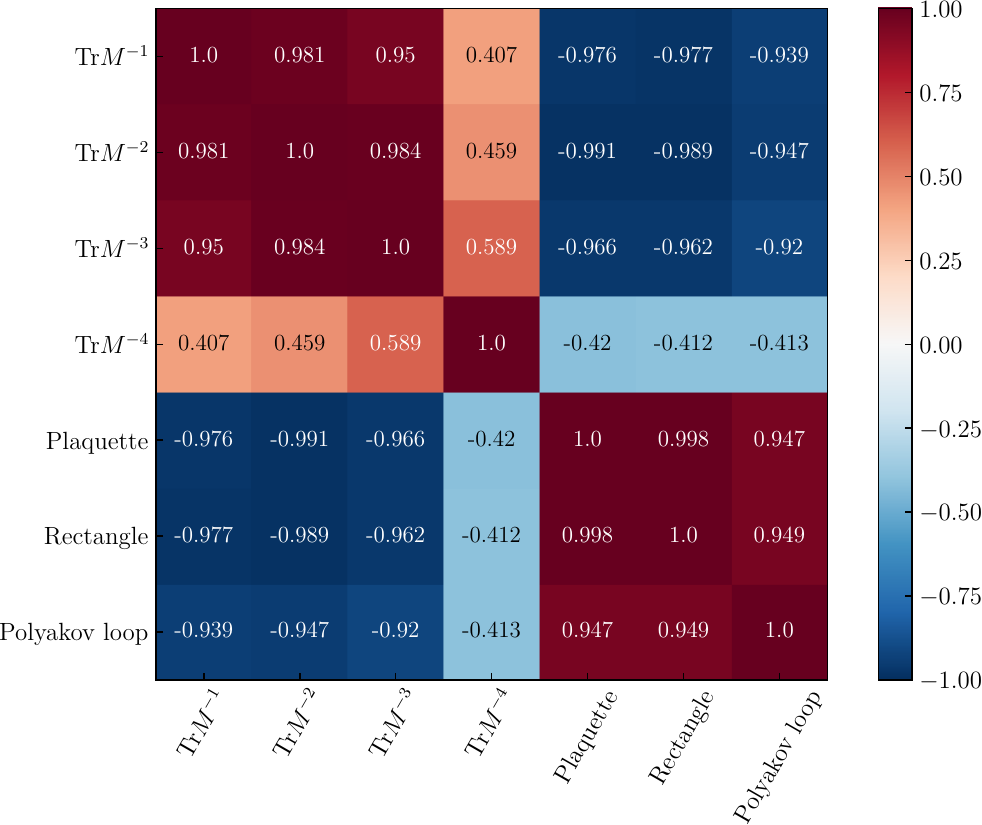}
        \label{fig:subfig-corr-ID-L12T4b1.60k13585}
  }
  \caption{Correlation between physical observables.}
  \label{fig:corr-obs-1}
\end{figure}  
%
Supervised regression benefits from strong correlations between input
features and target observables.
In Fig.~\ref{fig:corr-obs-1} we show the correlation matrix among
$\Tr\,M^{-n}$, the plaquette, the rectangle, and the Polyakov loop for
two representative ensembles, \texttt{L12T4b1.60k13575} (heaviest
quark) and \texttt{L12T4b1.60k13585} (lightest quark).
Apart from $\Tr\,M^{-4}$, which exhibits weaker correlations and thus
requires some care, most pairs display substantial positive
correlation, which motivates the use of these quantities as ML inputs
and targets.

\subsection{Two Practical ML Setups}
\label{sec:ML-approach}

The traces $\Tr\,M^{-n}$ entering cumulants are evaluated by a
Hutchinson-type stochastic trace estimator, in which random noise
vectors probe the operator, and the dominant cost arises from CG
inversions of the Dirac matrix.
To further reduce this cost, we consider two complementary choices of
input features:
\begin{description}[
    font=\bfseries,
    labelsep=1.2em,
    align=left,
  ]
  \ecitem{$\mathcal{F}_{\text{in}}$}{it:trace-feature}
  In this setup, the feature set $\mathcal{F}$ consists of
  \emph{internal} trace-based observables, in particular
  $\Tr\,M^{-1}$.
  We use the original CG measurements of $\Tr\,M^{-1}$ both as a
  direct input to the cumulant construction and as a feature to
  predict $\Tr\,M^{-n}$ ($n=2,3,4$).
  The ML-predicted higher powers are then combined with the exact
  $\Tr\,M^{-1}$ to build the cumulants.
  This configuration forms the main line of analysis in this work.
  Because $\Tr\,M^{-1}$ typically dominates the cumulant expressions,
  the resulting cumulants remain close to the original CG values even
  when $\mathcal{R}_{\text{LB}}$ is small; at the same time, the
  required CG work cannot be reduced below the cost of measuring
  $\Tr\,M^{-1}$.
  For cumulants up to kurtosis, this implies a lower bound of about
  $25\%$ of the baseline cost.
  \ecitem{$\mathcal{F}_{\text{ex}}$}{it:plaquette-feature}
  Here, the feature set $\mathcal{F}$ is built from \emph{external}
  observables recorded during the HMC evolution, namely the plaquette
  and rectangle, which enter the Iwasaki gauge action
  \cite{Iwasaki:1985we, Iwasaki:1983iya} and are available from the
  original simulations \cite{Ohno:2018gcx}.
  We use these two quantities as default input features and treat all
  $\Tr\,M^{-n}$ as ML targets, so that the quality of
  the cumulants fully reflects the ML estimation performance.
  Although the Polyakov loop is also measured, its correlation with
  $\Tr\,M^{-n}$ is weaker than that of the plaquette and rectangle
  (Fig.~\ref{fig:corr-obs-1}), and we therefore omit it from the
  default feature set.
\end{description}
The \ref{it:trace-feature} setup is closer to a ``partial
replacement'' strategy, where only higher powers of $M^{-1}$ are
delegated to ML, while the \ref{it:plaquette-feature} setup aims at a
fully feature-only prediction pipeline.
Comparing the two allows us to distinguish the impact of having
$\Tr\,M^{-1}$ explicitly available from the purely feature-based
scenario.

\subsection{Treatment of Statistical Errors}
\label{sec:stat-err}

The configurations analyzed here are generated near a first-order
phase transition and exhibit autocorrelations; naive delete-1
jackknife or \textit{i.i.d.}~bootstrap procedures that neglect this
dependence tend to underestimate statistical errors.
A more appropriate treatment uses either delete-$g$ jackknife or block
bootstrap resampling.

In our case, the ML estimator combines contributions from subsets of
different sizes (unlabeled and bias-correction sets); see
Eq.~\eqref{eq:P1-1}.
Constructing synchronized jackknife replicas across all these subsets
is cumbersome, so we adopt a block bootstrap approach to account for
temporal correlations in a straightforward manner.
Related discussions of block bootstrap methods in lattice QCD can be
found in Ref.~\cite{Christ:2024nxz}.

\subsection{Evaluation Criterion}

To quantify the agreement between the ML-based estimates and the
reference results obtained from the conventional CG method, we employ
a single scalar metric: the Bhattacharyya coefficient
\cite{Yoon:2018krb}.
For two Gaussian distributions, it is defined as
\begin{align}
  C_{\text{B}}(x,r) &= \sqrt{\frac{2r}{1+r^{2}}} \, \exp
  \left[-\frac{x^{2}}{4(1+r^{2})}\right] \,,
  \quad\text{where}\quad
  x = \frac{\left\lvert \bar{Y}_\text{Orig} - \bar{Y}_\text{ML}
    \right\rvert} {\sigma_\text{Orig}} \,, \quad r =
  \frac{\sigma_\text{ML}}{\sigma_\text{Orig}} \,.
  \label{eq:x-r-1}
\end{align}
Here, $x$ measures the normalized shift in the mean, and $r$
represents the relative size of the statistical uncertainty of the ML
estimation compared with the original result.

The coefficient $C_{\text{B}}$ takes values between $0$ and $1$, with
larger values corresponding to a stronger overlap between the two
distributions.
In the limiting cases where the ML estimator reproduces both the mean
($x=0$) and uncertainty ($r=1$) of the reference result,
$C_{\text{B}}=1$, indicating perfect statistical consistency.
Based on typical Gaussian overlap behavior, a value of $C_{\text{B}}
\gtrsim 0.95$ corresponds to a separation well below one standard
deviation and a relative uncertainty close to unity, and is therefore
interpreted here as indicating substantial agreement between the ML
prediction and the original measurement.
Throughout this work, $C_{\text{B}}$ serves as the primary figure of
merit, while $x$ and $r$ are referenced only when a more detailed
diagnostic is required.

\section{Cumulant estimation with multi-ensemble reweighting}
\label{sec:cumulant-reweighting}

Having established the ML setup and the feature choices, we now
proceed to apply the estimators to predict $\Tr\,M^{-n}$ on all
ensembles. These predictions are then combined through multi-ensemble
reweighting to extract the cumulants along the quark-mass trajectory.
In this setting, the trace estimates from all available
ensembles—generated at common $(V,\beta)$ but distinct values of
$\kappa$—are combined to interpolate observables across the quark-mass
axis following the standard Ferrenberg-Swendsen reweighting
framework~\cite{Ferrenberg:1988yz, Ferrenberg:1989ui}.
The underlying gauge ensembles were originally produced for
Ref.~\cite{Ohno:2018gcx} and have previously been used to investigate
the finite-temperature phase structure of $N_{\text{f}}=4$ QCD.

Our implementation combines the usual multi-ensemble machinery with
the bias-corrected ML estimators of Eq.~\eqref{eq:P1-1}.
For each observable $Y$ (here $Y$ represents the trace combinations
entering $Q_j$), the samples on every ensemble are partitioned into
$S^Y_{\text{TR}}$, $S^Y_{\text{BC}}$, $S^Y_{\text{UL}}$ and their
predicted counterparts $S^P_{\text{BC}}$, $S^P_{\text{UL}}$.
We then construct, for each ensemble, four index-aligned sets
\begin{align}
  \mathcal{S}_1 &= S^Y_{\text{TR}} \cup S^Y_{\text{BC}} \cup
  S^Y_{\text{UL}} \equiv S^Y \,, \qquad
  \mathcal{S}_2 = S^Y_{\text{TR}} \cup S^Y_{\text{BC}} \cup
  S^P_{\text{UL}} \,, \nonumber \\
  \mathcal{S}_3 &= S^Y_{\text{TR}} \cup S^Y_{\text{BC}} \equiv
  S^Y_{\text{LB}} \,, \qquad\qquad\;
  \mathcal{S}_4 = S^Y_{\text{TR}} \cup S^P_{\text{BC}} \,,
  \label{eq:EA-1-2-3-4}
\end{align}
ensuring that the configuration indices and ordering coincide with the
original data.
Afterward, the corresponding sets from all $\kappa$ values are
concatenated (\textit{e.g.}, in increasing $\kappa$) to form
cross-ensemble streams used in the subsequent reweighting steps.

\begin{figure}[tb]
  \vspace{-1.5em}
  \centering
  \subfigure[\ref{it:plaquette-feature} approach]{
    \includegraphics[width=0.48\linewidth]{
      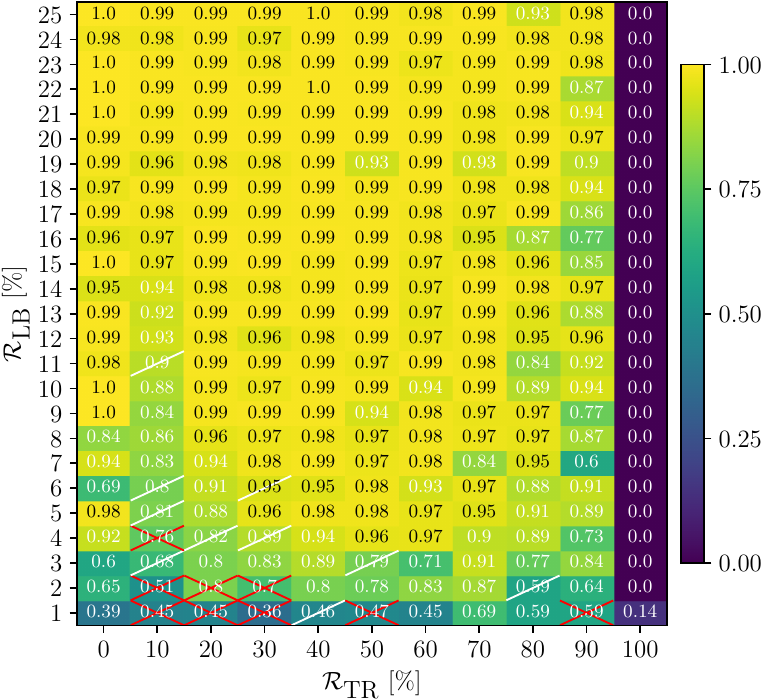
    }
    \label{fig:kurt-kurt-heatmap-ML2-LBP-1-25}
  }
  \hfill
  \subfigure[\ref{it:trace-feature} approach]{
    \includegraphics[width=0.48\linewidth]{
      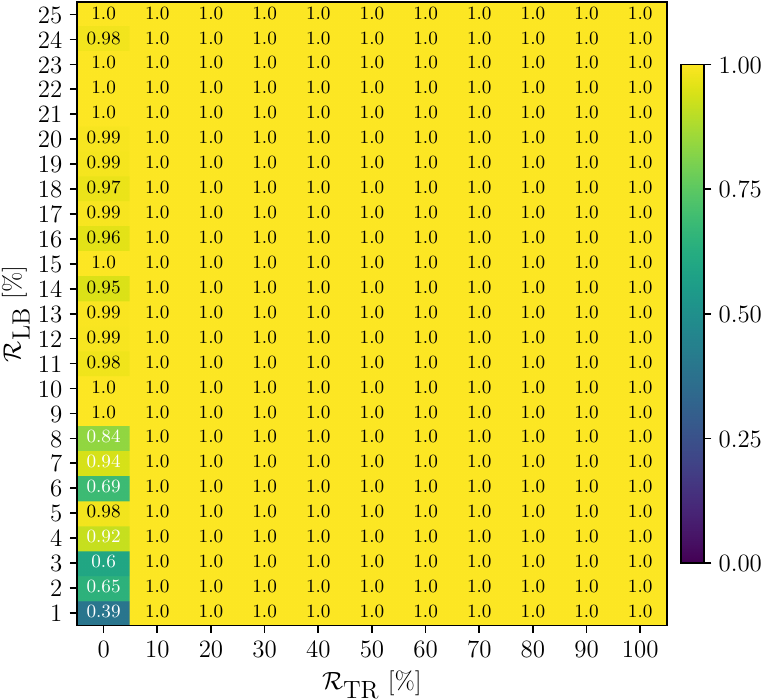
    }
    \label{fig:kurt-kurt-heatmap-ML1-LBP-1-25}
  }
  \caption{Bhattacharyya coefficient $C_\text{B}$ maps for the
    kurtosis at the transition point $K(\kappa_t)$ obtained from
    multi-ensemble reweighting.
    Panel~\subref{fig:kurt-kurt-heatmap-ML2-LBP-1-25} shows the
    results for $\mathcal{P}1$ under the \ref{it:plaquette-feature}
    setup.
    Red cross marks indicate cells where the Newton--Raphson solver
    failed to converge within the maximum iteration count, while white
    diagonal marks indicate cases where convergence was achieved but
    required more than $10$ Newton iterations.
    Panel~\subref{fig:kurt-kurt-heatmap-ML1-LBP-1-25} shows the
    corresponding $\mathcal{P}1$ results for the
    \ref{it:trace-feature} setup.}
  \label{fig:kurt-kurt-heatmap-LBP-1-25}
\end{figure}
%
We now discuss the multi-ensemble reweighting results for the
\ref{it:plaquette-feature} approach.
Before presenting the comparable cumulant outputs, we briefly note
that the Newton-Raphson solver used to determine the free-energy
offsets does not perform uniformly across the scanned
$\big(\mathcal{R}_{\text{LB}},\mathcal{R}_{\text{TR}}\big)$ parameter
space.
In particular, a small subset of cases with very limited labeled data
($\mathcal{R}_{\text{LB}}=1$--$4\%$ combined with low
$\mathcal{R}_{\text{TR}}$) failed to converge within the allowed
iteration limit, while some additional points did converge but only
after requiring more than $10$ iterations.
These behaviors were tracked in advance because they signal
instability of the ML-derived input traces rather than an issue with
the reweighting procedure itself, and the affected entries are
explicitly marked in the following figures.

With these caveats in place,
Fig.~\ref{fig:kurt-kurt-heatmap-LBP-1-25}\subref{fig:kurt-kurt-heatmap-ML2-LBP-1-25}
presents the Bhattacharyya coefficient $C_\text{B}$ for the kurtosis
estimation at the transition point $K(\kappa_t)$, obtained under the
\ref{it:plaquette-feature} setup using all ensembles in
Table~\ref{tab:ensemble-1}.

A clear pattern emerges.
First, in the low-$\mathcal{R}_{\text{LB}}$,
low-$\mathcal{R}_{\text{TR}}$ corner the Newton solver is unreliable
and the resulting $C_\text{B}$ is also poor.
Even nearby cells without explicit convergence issues tend to show
suppressed overlap, reflecting the limited quality of
$S^P_{\text{UL}}$ when the labeled fraction is too small.
For this dataset the problematic region largely disappears once
$\mathcal{R}_{\text{LB}}\gtrsim 20\%$.
Second, the combination of small $\mathcal{R}_{\text{LB}}$ and large
$\mathcal{R}_{\text{TR}}$ (\textit{i.e.}~very small
$\mathcal{R}_{\text{BC}}$) again leads to reduced $C_\text{B}$.
In particular, the column $\mathcal{R}_{\text{TR}}=90\%$ only reaches
stable high-overlap values once $\mathcal{R}_{\text{LB}}\gtrsim 23\%$,
similar to the behavior at $\mathcal{R}_{\text{TR}}=10\%$.
This indicates that the ML quality degrades both when the training set
is extremely small and when the bias-correction set becomes too small.

\begin{figure}[tb]
  \vspace{-1.2em}
  \includegraphics[width=0.97\linewidth]{
    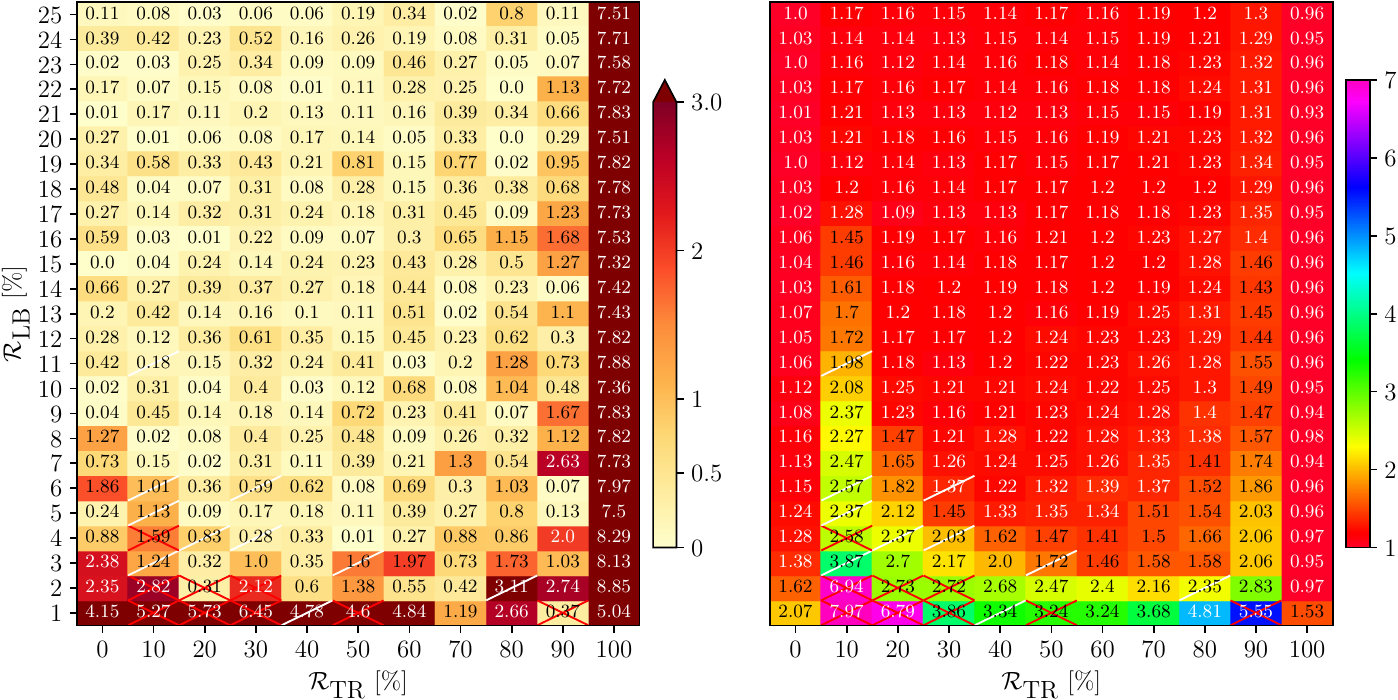
  }
  \caption{ Heatmap showing the $\mathcal{P}1$ results of $x$ and $r$
    evaluations, where $x$ and $r$ are defined in
    Eq.~\eqref{eq:x-r-1}, for the estimation of the kurtosis at the
    phase transition point, $K(\kappa_t)$, based on the
    \ref{it:plaquette-feature} approach.
    Within the panel, the left half represents the $x$ results, while
    the right half corresponds to $r$.
    In the $x$ maps, lighter shades indicate smaller normalized mean
    separations $x$, corresponding to better agreement between the
    original and predicted means, whereas darker shades represent
    larger $x$ values, signaling poorer overlap.
    The kurtosis is computed using $\Tr\,M^{-n}$
    predicted from plaquette and rectangle as the input feature. }
  \label{fig:kurt-kurt-heatmap-ML2-LBP-1-25-EC1-EC2}
\end{figure}
%
The most striking feature appears in the
$\mathcal{R}_{\text{TR}}=100\%$ column, where the labeled data are
used exclusively for training and no bias correction is performed.
Here $C_\text{B}$ is essentially zero throughout the scan.
The $x$ (see Eq.~\eqref{eq:x-r-1}) analysis in
Fig.~\ref{fig:kurt-kurt-heatmap-ML2-LBP-1-25-EC1-EC2} shows that the
normalized mean separation reaches $x\simeq 7$–$8$, indicating that
the ML result for $K(\kappa_t)$ deviates by several standard
deviations from the conventional reweighting result.
Similar behavior, though with varying magnitude, is seen in other
datasets (not shown), systematically confirming that removing the bias
correction at the reweighting stage leads to significant distortions
once higher-order cumulants and the transition point are inferred.

For comparison,
Fig.~\ref{fig:kurt-kurt-heatmap-LBP-1-25}\subref{fig:kurt-kurt-heatmap-ML1-LBP-1-25}
displays the corresponding multi-ensemble reweighting results for the
\ref{it:trace-feature} approach.
In this case the $\mathcal{P}1$ estimator yields $C_\text{B}\approx 1$
essentially everywhere in the scanned
$\big(\mathcal{R}_{\text{LB}},\mathcal{R}_{\text{TR}}\big)$ range,
except for the low-$\mathcal{R}_{\text{LB}}$ region at
$\mathcal{R}_{\text{TR}}=0\%$, where only the labeled set is used and
the statistical precision is inevitably worse than for the full
original data.
Thus, when the exact $\Tr \, M^{-1}$ is retained as in the
\ref{it:trace-feature} setup, the reweighting-based cumulants and the
extracted transition point remain remarkably robust against changes in
$(\mathcal{R}_{\text{LB}},\mathcal{R}_{\text{TR}})$, whereas in the
fully ML-driven \ref{it:plaquette-feature} case, bias correction plays
a crucial role in stabilizing the multi-ensemble cumulant analysis.

\vspace{-0.1em}

\section{Conclusion}
\label{sec:conc}

In this work we have applied the bias-corrected ML framework of
Ref.~\cite{Yoon:2018krb} to the estimation of thermodynamic
observables relevant to finite-temperature QCD, and compared the
results with conventional calculations performed on the full set of
stochastic trace measurements.

For the \ref{it:trace-feature} approach, the ML-assisted estimates
reproduce the original results for traces, single-ensemble cumulants,
and multi-ensemble reweighting almost perfectly across all tested
datasets and parameter pairs, including the most extreme case with
$\mathcal{R}_{\text{LB}}=1\%$.
This robustness is naturally understood from the fact that the full
set of measured $\Tr M^{-1}$ values is kept, and the cumulants are
dominated by this observable.
Under this assumption, the total cost can, in our setup, be reduced to
about
\[
\frac{100+1+1+1}{400} \simeq 25.75\%
\]
of the original measurement budget while retaining the same precision,
suggesting a sizable practical gain if the observed stability
persists in broader applications.

The \ref{it:plaquette-feature} approach is more ambitious: all $\Tr \,
M^{-n}$'s are inferred from gauge observables instead of
lower-order traces.
Consequently, the overlap with the full-data results, quantified by
the Bhattacharyya coefficient $C_{\text{B}}$, shows a stronger
dependence on the size of the labeled set.
In particular, for the ensembles studied here, $C_{\text{B}}$ improves
systematically with increasing $\mathcal{R}_{\text{LB}}$ and becomes
more stable once $\mathcal{R}_{\text{LB}}\gtrsim 20\%$, indicating a
potential cost reduction to the $\sim 20\%$ level.
At the same time, the use of independent gauge features makes this
setup more sensitive to modeling choices, and careful validation
over additional ensembles and possibly more sophisticated ML
architectures will be needed before drawing firm conclusions.

A lesson from the \ref{it:plaquette-feature} analysis is the
importance of bias correction, especially when ML outputs are fed into
multi-stage workflows.
In the extreme case $\mathcal{R}_{\text{TR}}=100\%$, where the labeled
data are used only for training and no bias correction is performed,
the agreement with the full-data results deteriorates markedly as one
proceeds from trace estimation to multi-ensemble reweighting and
finally to the determination of the kurtosis at the transition point
$\kappa_t$.
The accumulated shift in this ``no-bias-correction'' column reaches
many standard deviations, underscoring that even small residual biases
in intermediate ML predictions can be amplified through higher-order
observables and interpolation procedures.

In summary, the \ref{it:trace-feature} strategy appears to be a
practical and robust option for reducing the computational footprint
of fermionic observables, at least for the ensembles and observables
considered in this study.
The \ref{it:plaquette-feature} strategy is more challenging but also
potentially more flexible, and our results indicate that its reliable
use will require both a sufficient labeled fraction and explicit bias
correction at the cumulant level.

We expect that further tests on different lattices and actions, and
extensions to other thermodynamic and fluctuation observables, will
clarify how broadly this ML-based framework can be deployed within the
lattice QCD community.

\acknowledgments

B.~J.~C. would like to thank Takayuki Sumimoto for his early
contributions and dedication to the initial stage of this work.
He also thanks Ho Hsiao for fruitful discussions.
The work of A.~T. was partially supported by JSPS KAKENHI Grants
No.~20K14479, No.~22H05111, No.~22K03539 and JST BOOST, Japan Grant
No.~JPMJBY24F1.
A.~T. and H.~O. were partially supported by JSPS KAKENHI
Grant No.~22H05112.
B.~J.~C. and part of this work were supported by MEXT as ``Program for
Promoting Researches on the Supercomputer Fugaku'' (Grant Number
JPMXP1020230411, JPMXP1020230409).


\bibliography{ref}

\end{document}

%% file: macro.tex
\newcommand{\Tr}{\mathrm{Tr}}

\newcolumntype{\.}{>{\global\let\currentrowstyle\relax}}

\newcolumntype{^}{>{\currentrowstyle}}

\makeatletter
\newcommand{\ecitem}[2]{%
  \item[\textbf{#1}]%
  \phantomsection
  \protected@edef\@currentlabel{\unexpanded{\textbf{#1}}}%
  \label{#2}%
}
\makeatother

\newcommand{\orcidauthorOHNO}{0000-0003-1798-8222}
\newcommand{\orcidauthorTOMIYA}{0000-0001-9374-3716}
\newcommand{\orcidauthorCHOI}{0000-0002-5438-5490}